\documentclass[prd, aps, showkeys, twocolumn, superscriptaddress, showpacs, nofootinbib, usenatbib, longbibliography]{revtex4-2}

\usepackage{graphicx}% Include figure files
\usepackage{dcolumn}% Align table columns on decimal point
\usepackage{bm}% bold math
\usepackage{lipsum}
\usepackage{amsmath,amssymb}
\usepackage[bottom]{footmisc}
\usepackage{booktabs}
\usepackage[utf8]{inputenc}
\usepackage[T1]{fontenc}
\usepackage[normalem]{ulem} %% For striketrhough command
\usepackage{verbatim}
\usepackage[left=2cm, right=2cm, top=2cm]{geometry}

\usepackage[]{algorithm2e}
\usepackage[framemethod=tikz]{mdframed}
\usepackage{epigraph}
\usepackage{siunitx}
\usepackage[english]{babel}
\usepackage{listings}
\usepackage{graphicx}
\usepackage[utf8]{inputenc}
\usepackage{listings}
\usepackage{matlab-prettifier}
\usepackage{amsfonts}
\usepackage{amsmath}
\usepackage{amssymb}
\usepackage{amsthm}
\usepackage{array}
\newcolumntype{P}[1]{>{\centering\arraybackslash}p{#1}}
\newcolumntype{M}[1]{>{\centering\arraybackslash}m{#1}}
\usepackage{braket}
\usepackage{xcolor}
\usepackage{enumitem}   
\usepackage{tabularx}
\usepackage{booktabs}
\usepackage{mathtools}

\usepackage{multirow}

\usepackage{glossaries}

\usepackage[colorlinks = true,
            linkcolor = blue,
            urlcolor  = blue,
            citecolor = blue,
            anchorcolor = blue]{hyperref}
\usepackage{comment}
\usepackage{color}

\usepackage[capitalise]{cleveref}
\usepackage[commandnameprefix=ifneeded]{changes}

\usepackage{orcidlink}

\graphicspath{{figures/}}

\graphicspath{ {./figures/} }

\begin{document}

\preprint{APS/123-QED}

\title{First-time assessment of glitch-induced bias and uncertainty in inference of extreme mass ratio inspirals}% Force line breaks with \\

%Potential titles:

%Glitch-induced biases from parameter estimation of extreme mass ratio inspirals

\author{Amin Boumerdassi\, \orcidlink{0000-0002-8260-4072}}
 \email{abou623@aucklanduni.ac.nz}
\author{Matthew C. Edwards\, \orcidlink{0000-0003-2861-5616}}%
 % \email{matt.edwards@auckland.ac.nz}
\author{Avi Vajpeyi\, \orcidlink{0000-0002-4146-1132}}%
 % \email{avi.vajpeyi@auckland.ac.nz}
\affiliation{Department of Statistics, University of Auckland, New Zealand}%
\author{Ollie Burke\,\orcidlink{0000-0003-2393-209X}} 
\affiliation{School of Physics and Astronomy, University of Glasgow, Glasgow G12 8QQ, UK}
\affiliation{Laboratoire des 2 Infinis - Toulouse (L2IT-IN2P3), Université de Toulouse, CNRS, F-31062 Toulouse Cedex 9, France}
  % \email{Ollie.Burke@glasgow.ac.uk}

\date{\today}% It is always \today, today,

\begin{abstract}
This work investigates the impact of streams of transient, non-Gaussian noise artifacts or ``glitches'' on the parameter estimation of extreme mass ratio inspirals (EMRI) in the Laser Interferometer Space Antenna (LISA). 
Glitches cause biased and less precise inference for short-duration signals such as massive black hole binaries, but their effect on long-lived sources such as EMRIs has not been quantified. 
Using simulated LISA observations containing injected EMRIs and streams of shapelet-based glitches drawn from the LISA Pathfinder catalog, we estimate the glitch-induced parameter biases and uncertainties through a Fisher-matrix-based analysis whose accuracy we verify with Markov-Chain Monte Carlo.
We find that moderately mitigated glitch streams i.e. ones containing only glitches of up to moderate SNRs ($\rho \lesssim 90$) induce negligible to minor biases $[\sim0.04\sigma ,\sim0.6\sigma]$ in the inferred EMRI parameters. In contrast, weakly mitigated glitch streams containing higher-SNR events ($\rho \lesssim 400$) can produce biases nearing $1\sigma$. 
These results demonstrate that, when compared to inference of other sources such as massive black hole binaries, EMRI inference is notably more robust to glitches.
We stress that at least some amount of glitch modeling and mitigation remains essential for unbiased EMRI analyses in the LISA era.
\end{abstract}

\maketitle

\section{Introduction}

The launch of the Laser Interferometer Space Antenna (LISA)~\cite{DANZMANN20001129} in the mid-2030s will open access to the millihertz gravitational-wave (GW) window, enabling observations in the frequency band $[10^{-4},10^{-1}]\,\mathrm{Hz}$ that bridges those accessible to ground-based detectors and pulsar timing arrays. 
During its 4.5-year mission, LISA is expected to detect up to $\sim10^3$ events from extreme mass ratio inspirals (EMRIs)~\cite{Gair_2017,Babak_2017,Babak_2010,babak2017sciencespacelisa}, as well as numerous galactic binaries, massive black hole mergers, stochastic backgrounds, stellar origin black hole binaries and, potentially, exotic sources~\cite{colpi2024lisadefinitionstudyreport}.

Of particular interest to the LISA community are EMRIs which occur when a stellar-mass compact object of mass $\mu\in[10,10^2]\,M_\odot$ spirals into a supermassive black hole of mass $M\in[10^4,10^7]\,M_\odot$. 
Due to extreme mass ratios of $\eta = \mu/M \lesssim 10^{-4}$, these systems are characterized by long-lived and highly relativistic waveforms that last for tens to hundreds of thousands of orbital cycles, and durations of months to years. 
LISA will be sensitive to such signals with optimal signal-to-noise ratios (SNRs) in the range $[20,100]$~\cite{babak2017sciencespacelisa}.

Parameter estimation (PE) for EMRIs promises unprecedented precision and scientific return~\cite{berry2019uniquepotentialextrememassratio}. For typical signals, uncertainties in quantities such as the eccentricity and component masses may reach the $\mathcal{O}(10^{-4})$ level, enabling precision tests of general relativity (GR) in the strong-field regime~\cite{PhysRevD.75.042003,Gair_2013}, independent constraints on cosmological parameters~\cite{laghi2021gravitationalwavecosmologyemris,toscani2023stronglylensedextrememassratioinspirals}, and searches for deviations from the Kerr spacetime~\cite{kumar2024probingdeviationskerrgeometry,kumar2024gravitationalwavesregularblack,barsanti2023detectingmassive,speri2024probingfundamental,maselli2022detectingfundamental}. 
Such extraordinary accuracy and precision may in practice be degraded by model mis-specification in the form of waveform mismodelling~\cite{khalvati2025systematicerrorsfastrelativistic}, astrophysical environments~\cite{Speri_2023} broken assumptions on noise properties~\cite{Burke:2025bun} etc.

Consequently, EMRI PE demands accurate and computationally efficient waveform models~\cite{burke2024accuracyrequirementsassessingimportance,PhysRevD.96.044005,GAIR_2008,Chua_2015}. 
Existing approaches build on black hole perturbation and self-force theory~\cite{10.1143/PTPS.128.1,PhysRevD.109.044020,PhysRevD.106.104025,Barack_2018,PhysRevD.104.084011,PhysRevD.106.104001}, while recent advances leverage reduced-order surrogates, GPU acceleration, and machine-learning techniques~\cite{Chua:2020stf,Katz:2021yft,michael_l_katz_2023_8190418,Chua:2018woh,Chua:2017ujo,Speri:2023jte,katz2021fastemriwaveforms,khalvati2024impactrelativistic}, and references therein. 

An important consideration in EMRI data analysis is the statistical properties of the detector noise.
While most LISA analyses assume stationary and Gaussian noise, real GW data contain non-stationary, non-Gaussian transient events known as \emph{glitches}.
If unaccounted for, these events can result in biased and imprecise PE.
The LISA Pathfinder mission~\cite{mcnamara2006lisa,Antonucci_2012,Armano_2015} directly characterized LISA-like glitches, recording nearly 600 such events during its two-year operation~\cite{PhysRevD.110.042004,baghi2022detectioncharacterization}.
These glitches occurred on average once per day, and are believed to originate from outgassing and other space-environment effects~\cite{PhysRevD.106.062001,Sala:2023hpr}.

A wide range of glitch detection and mitigation strategies have been proposed such as data masking, hybrid neural-network classification~\cite{Houba_2024,houba2024detectionmitigation}, matching-pursuit decompositions~\cite{PhysRevD.105.042002}, reversible-jump Markov chain Monte Carlo methods~\cite{muratore2025pipelinesearchingfittinginstrumental}, and wavelet-based analyses~\cite{licciardi2025waveletscatteringtransformgravitational}.
Time–frequency representations~\cite{gair2007timefrequency,hughes2023fastfourier} and global-fitting pipelines~\cite{katz2024efficientgpu,littenberg2023prototypeglobal,deng2025modularglobal,strub2024globalanalysis} have further improved the ability to separate overlapping sources and glitches.
Recent LISA data challenges have incorporated realistic glitch populations derived from Pathfinder observations~\cite{baghi2022lisadatachallenges}.

Despite these developments, their implications for EMRI inference remain uncertain.
The long duration and complex evolution of EMRIs make it unclear when and how many glitches would induce measurable biases.
Previous studies have examined single-glitch biases for shorter-duration sources such as massive black hole binaries and galactic binaries~\cite{Spadaro_2023,castelli2024extractiongravitationalwavesignals}, but no analogous investigation has yet been conducted for EMRIs.

In this work, we assess the impact of streams of glitches—hereafter referred to as \emph{glitch backgrounds}—on EMRI PE.
We employ a Fisher-matrix-based analysis to estimate the inference bias and uncertainty due to glitch-related model misspecification.
By considering glitch backgrounds mitigated to varying degrees, we aim to establish preliminary constraints on how lax the glitch mitigation can be before observing significant biases thereby addressing a critical and previously unexplored gap in LISA data analysis preparedness.

The remainder of this paper is organized as follows. Section~\ref{sec:data_analysis_fundamentals} outlines the GW data analysis fundamentals, including the Fisher-matrix formalism, and biases from glitch backgrounds. Section~\ref{sec:modelling} covers the EMRI waveform and glitch modeling used in this paper, as well as our agnostic method for generating ``mitigated'' glitch backgrounds. 
Section~\ref{sec:results} presents the main findings, quantifying the relationship between glitch SNR and parameter bias. Section~\ref{sec:conclusions} discusses the implications of our findings for future LISA analyses and summarizes our conclusions and prospects for future work.  Before presenting our methodology, we briefly summarize the conventions and notations used throughout this work in Section~\ref{sec:conventions}.

\subsection{Conventions}\label{sec:conventions}

Time-domain quantities are represented by discrete vectors $a(t_{i}) \approx \boldsymbol{a}[i]$ for $i \in \{0,1,\ldots,N-1\}$, sampled at times $t_i = i\,\Delta t$, with sampling interval $\Delta t = t_{i+1}-t_i$ and total observation time $t_N = N\Delta t = T_{\mathrm{obs}}$.  
Analogous quantities in the frequency domain are denoted by $\boldsymbol{\tilde{a}}(f_i)$, sampled at discrete Fourier frequencies $f_i = i\,\Delta f$, where $\Delta f = 1/(N\Delta t) = 1/T_{\mathrm{obs}}$.  
The Fourier-domain vectors are sampled at rate $f_s = 1/\Delta t$ over positive frequencies $f_i \in \{0,\Delta f,\ldots,f_s/2\}$. Tilded quantities $\boldsymbol{\tilde{a}}$ represent the Fourier transform of time-domain expressions. 
We define the continuous Fourier transform as
\begin{equation}
    \tilde{a}(f) = \int_{-\infty}^{\infty} a(t)\, e^{-2\pi i f t}\, \mathrm{d}t,
\end{equation}
and its discrete analog as
\begin{equation}
    \tilde{a}(f_j) \approx \tilde{a}[j] = \Delta t \sum_{k=0}^{\lfloor N/2 + 1\rfloor} a[k]\, 
    e^{\,-2\pi i (kj)/N},
\end{equation}
whose components form the discrete vector $\boldsymbol{\tilde{a}}_j = \tilde{a}_j$ for $j \in \{0,1,\ldots,\lfloor N/2 + 1\rfloor\}$.

Parameter vectors are denoted in boldface, $\boldsymbol{\theta} = \{\theta^i\}$, with derivatives written compactly as $\partial_i a \equiv \partial a / \partial \theta^i$.  
For stochastic processes, we write the covariance as $\mathbb{E}[\boldsymbol{a}\boldsymbol{a}^\dagger] = \boldsymbol{\Sigma}$, where $\boldsymbol{a}^\dagger$ is the complex conjugate transpose of $\boldsymbol{a}$ and $\mathbb{E}$ denotes the expectation over the data-generating process.  
For ergodic, infinite-duration processes, the ensemble average $\langle \cdot \rangle$ replaces $\mathbb{E}$ and represents an average over the observation period.  
Unless stated otherwise, we adopt the Einstein summation convention: repeated upper and lower indices are summed over their full range, e.g., $\boldsymbol{a}_{ij}\boldsymbol{b}_j = \sum_j \boldsymbol{a}_{ij}\boldsymbol{b}_j$.

\section{Estimating Parameters, Errors and Biases}\label{sec:data_analysis_fundamentals}

\subsection{Data-stream and likelihood model}
\label{The data-stream and likelihood}
We simulate time-domain GW data $d$ as the sum of the EMRI's GW strain $h$ parameterized by the true astrophysical parameters ${\boldsymbol{\theta}}_0$, a zero-mean stationary Gaussian noise process $n$, and a non-Gaussian non-stationary glitch background $g$ which is a stream of $N$ independent and potentially overlapping glitches parameterized by vector $\boldsymbol{\Lambda}$ which contains each individual glitch's set of parameters $\boldsymbol{\lambda}^{(i)}$.
Our glitch stream has the form
\begin{equation}\label{eq:glitch_process_sum}
    g(t;\boldsymbol{\Lambda}) = \sum_{i=1}^{N} g(t;\boldsymbol{\lambda}^{(i)}),
\end{equation}
and so the data can be expressed as
\begin{equation}
\label{eq:GW_model_eqn}
    d(t) = h(t;\boldsymbol{\theta}_{0})+ g(t;\boldsymbol{\Lambda}) + n(t)\,.
\end{equation}

Throughout this work, we decouple the stationary noise process from the non-stationary transient process.
The stationary noise is entirely characterized by its one-sided Power Spectral Density (PSD) $S_{n}(f)$, defined through 
\begin{subequations}\label{eq:stat_noise_covariances}
\begin{align}
\langle\tilde{n}(f)\tilde{n}^{\star}(f')\rangle &= \tfrac{1}{2}\delta(f - f')S_{n}(f')\,, \label{eq:covariance_noise_stat_continuous} 
\end{align}
\end{subequations}
implying that the frequency-domain noise components are uncorrelated. 
Independent noise realizations can be generated via standard frequency-domain sampling techniques such as those discussed in Appendix~B of~\cite{Burke:2025bun} and references therein.

The glitch stream $g(t;\boldsymbol{\Lambda})$ is a stochastic process with statistical properties informed by the population of test-mass interferometer (TMI) glitches observed in LISA Pathfinder data~\cite{PhysRevD.106.062001}.
For the interested reader, we elaborate more on the impact of such glitch backgrounds on PE in Section~\ref{Sec:Biases from glitch backgrounds}, while specific details of the glitch modeling and statistics are discussed in Sections~\ref{sec:modelling} and~\ref{sec:results} respectively.

We define the standard noise-weighted inner product between time series $a(t)$ and $b(t)$ as 
\begin{align}\label{eq:inner_product}
    (a|b) &= 4\mathrm{Re}\int^{\infty}_{0}\frac{\tilde{a}(f)\,\tilde{b}^*(f)}{S_{n}(f)}\,\mathrm{d}f\,, \\
& \approx 2\text{Re}\left(\tilde{\boldsymbol{a}}^{\dagger}\boldsymbol{\Sigma}^{-1}\tilde{\boldsymbol{b}}\right)\,, \label{eq:discrete_inner_product}
\end{align}
where the discrete noise covariance matrix $\boldsymbol{\Sigma}$ has diagonal components obtained from Eq.\eqref{eq:covariance_noise_stat_continuous}: 
\begin{equation}\label{eq:diagonal_noise_cov}
(\mathbb{E}[\boldsymbol{\tilde{n}}\boldsymbol{\tilde{n}}^{\dagger}])_{ij} = \boldsymbol{\Sigma}_{ij} = \frac{S_{n}(f_{i})\delta_{ij}}{2\Delta f}\,.
\end{equation}
The optimal matched-filtering SNR is then
\begin{equation}\label{eq:optimal_SNR}
    \rho_{\mathrm{opt}}=\sqrt{(h|h)}\,.
\end{equation}

Our goal is to quantify the inference bias and uncertainty that arise from glitch-related model misspecification.
By adopting the conventional Whittle-likelihood~\cite{whittle1951hypothesis, Finn_1992} while injecting glitches into the data, we break the assumption of purely stationary and Gaussian noise.
The logarithmic Whittle likelihood has the form
\begin{subequations}\label{eq:whittle_likelihood_both}
\begin{align}
    \log \mathcal{L}(d|\boldsymbol{\theta}) &= -\tfrac{1}{2}(d-h_{\rm m}(\boldsymbol{\theta})|d-h_{\rm m}(\boldsymbol{\theta}))\,,\label{eq:whittle_likelihood_continiuous} \\
    & \approx -(\tilde{\boldsymbol{d}} - \tilde{\boldsymbol{h}}_{\rm m}(\boldsymbol{\theta}))^{\dagger}\boldsymbol{\Sigma}^{-1}(\tilde{\boldsymbol{d}} - \tilde{\boldsymbol{h}}_{\rm m}(\boldsymbol{\theta}))\,.\label{eq:whittle_likelihood_discrete}
\end{align}
\end{subequations}
In Eq.~\eqref{eq:whittle_likelihood_both}, the data $d$ are described by Eq.~\eqref{eq:GW_model_eqn}, and $h_{\rm m}$ denotes the waveform model. We define the best-fit parameters $\widehat{\boldsymbol{\theta}}$ to be the parameters where the Eq.~\eqref{eq:whittle_likelihood_both} is maximized so as to satisfy
\begin{equation}
\partial_{i}(h_{m}(\boldsymbol{\theta})|d - h_{m}(\boldsymbol{\theta}))|_{\boldsymbol{\theta} = \hat{\boldsymbol{\theta}}} = 0\,.
\end{equation}

We stress that we do not attempt any glitch fitting.
Alternative approaches, such as the reversible-jump Markov-chain Monte-Carlo framework proposed in~\cite{muratore2025pipelinesearchingfittinginstrumental} include parameterized glitch models directly in the likelihood.
Such a joint signal-glitch inference for EMRIs is beyond the scope of this work, and left for future studies.

GW inference usually uses stochastic techniques such as Markov-chain Monte Carlo (MCMC), nested sampling or other global optimizers~\cite{chua_2022, 10.1093/mnras/stad2939, bilby:2019, dynesty:2020}. We note that EMRI likelihood surfaces are highly multimodal and so traditional samplers often become trapped in local maxima, yielding parameter estimates far from the global posterior mode~\cite{chua_2022}. The greatest advancement on the EMRI search can be found in Ref.~\cite{speri2025abunodisceomnes}, which details an efficient search pipeline for a single non-rotating but moderately eccentric (plunging) EMRI embedded in one year's worth of Gaussian and stationary noise at SNR $\sim 30$. For our analysis, the focus of our work is \emph{not} to search for the parameters, but understand the impact on parameter estimation of EMRIs in the presence of noise transients. For this reason, we will utilize tight priors around the injected EMRI parameters and use starting coordinates (that initialise our samplers) very close to the true parameters of the source.  

We also note that EMRI PE is computationally intensive and time-consuming, even with GPU-accelerated sampling and likelihoods.
Obtaining well-sampled posteriors typically requires hundreds of thousands of likelihood calls requiring over tens of hours of walltime.
In this work, we inject single EMRIs into glitch backgrounds mitigated to varying levels to quantify the resulting parameter bias and uncertainty.
The inclusion of glitches introduces an additional source of stochasticity in the inference, and so the bias estimation requires averaging glitch-induced fluctuations in the maximum likelihood estimate over many inference runs.
For EMRIs, this approach is computationally time-consuming, and so we instead leverage the Fisher-matrix formalism to cheaply and rapidly estimate the maximum likelihood estimates. 
\subsection{The Fisher-matrix formalism}
\label{sec:fisher_matrix}

To estimate parameter uncertainties and maximum likelihood estimates (MLE), we employ the Fisher-matrix (FM) formalism~\cite{Finn_1992}.  
In the high-SNR limit and around the likelihood maximum, the log-likelihood is a multivariate Gaussian distribution whose curvature defines the FM and sets the Cramér–Rao lower bound on achievable precisions of parameter estimates.

For small deviations from the true parameters, $\theta^i = \theta_0^i + \Delta\theta^i$ with $|\Delta\theta^i|\ll1$, the linear signal approximation (LSA) expands the model waveform as
\begin{equation}\label{eq:LSA}
h_{\mathrm{m}}(t;\boldsymbol{\theta})
\simeq
h_{\mathrm{m}}(t;\boldsymbol{\theta}_0)
+ \partial_j h_{\mathrm{m}}(t;\boldsymbol{\theta}_0)\,\Delta\theta^j.
\end{equation}
Substituting this into the Whittle likelihood in Eq.~\eqref{eq:whittle_likelihood_both}, and using the data model in Eq.~\eqref{eq:GW_model_eqn} (with $h_{\mathrm{m}} = h$) yields a quadratic form in $\Delta\theta^i$,
\begin{equation}\label{eq:fisher_likelihood}
\log\mathcal{L}
\propto
-\tfrac{1}{2}
(\Delta\theta^i-\widehat{\Delta\theta^i})
\,\Gamma_{ij}\,
(\Delta\theta^j-\widehat{\Delta\theta^j})\,,
\end{equation}
where the FM is defined as
\begin{equation}\label{eq:Fisher_Matrix_Components}
\Gamma_{ij} = (\partial_i h|\partial_j h).
\end{equation}
with Eq.\eqref{eq:fisher_likelihood} maximized at the value 
\begin{equation}\label{eq:MLE_total}
\widehat{\Delta\theta^i} = (\Gamma^{-1})^{ij}(\partial_j h|n+g)\,,
\end{equation}
and both the FM and MLEs are evaluated at the true parameters $\boldsymbol{\theta}=\boldsymbol{\theta}_0$.
We also note that numerical computation of the FM is notoriously ill-conditioned for EMRIs.
For this, we utilize the \texttt{StableEMRIFisher} package~\cite{kejriwal_2024_sef}, and verify its accuracy with the \texttt{eryn} sampler~\cite{Karnesis_2023} as described in Appendix~\ref{app_sec:FM_validation}.
The total $\widehat{\Delta\theta^i}$ in Eq.~\eqref{eq:MLE_total} separates naturally into independent contributions from stationary Gaussian noise and from transient glitches:
\begin{align}\label{eq:sum_of_MLEs_glitches_noise}
\widehat{\Delta\theta^i}_{\mathrm{noise}} &= (\Gamma^{-1})^{ij}(\partial_j h|n), \\
\widehat{\Delta\theta^i}_{\mathrm{glitches}} &= (\Gamma^{-1})^{ij}(\partial_j h|g).
\end{align}
This follows the Cutler–Vallisneri formalism~\cite{PhysRevD.57.4566,PhysRevD.71.104016,PhysRevD.76.104018,Vallisneri_2008,Antonelli_2021} where the glitch background $g$ replaces the confusion noise contribution studied in Ref.~\cite{Antonelli_2021}.

We know a priori that in the absence of glitches, the MLEs are normally distributed around the true parameters with covariance equal to the inverse of the FM.
This implies that
\begin{equation}
    \mathbb{E}[\widehat{\Delta\theta^i}_{\mathrm{noise}}]=0\implies\text{unbiased}\,,
\end{equation}
and
\begin{equation}\label{eq:covariance_noise_stat}
\mathrm{Cov}(\Delta\theta^i_{\mathrm{noise}},\Delta\theta^j_{\mathrm{noise}})=(\Gamma^{-1})^{ij}\,,
\end{equation}
meaning that
\begin{equation}
(\widehat{\theta^i}
=
\theta_0^i+\widehat{\Delta\theta^i}_{\mathrm{noise}})\sim\mathcal{N}(\theta_0^i,(\Gamma^{-1})^{ii})\,.
\end{equation}

For comparison, waveform-systematic studies often define the ratio~\cite{Vallisneri_2008,PhysRevD.76.104018,PhysRevD.57.4566,PhysRevD.71.104016,Antonelli_2021}
\begin{align}\label{eq:R_statistic_equations}
\mathcal{R}^i_{\mathrm{sys}} &=
\bigg(\frac{\Delta\theta^i_{\mathrm{sys}}}{\Delta\theta^i_{\mathrm{stat}}}\bigg), \\
\Delta\theta^i_{\mathrm{sys}} &= (\Gamma^{-1})^{ij}(\partial_j h|h_{\mathrm{e}}-h_{\mathrm{m}})\,,\\
\Delta\theta^{i}_{\text{stat}} &= \sqrt{(\Gamma^{-1})^{ii}}\,\quad \text{no sum}\label{eq:stat_uncertainty}
\end{align}
with $|\mathcal{R}|>1$ indicating that deterministic waveform biases exceed statistical fluctuations.  
In our case, the stochastic nature of the glitch background means that the corresponding MLE is also stochastic thereby being characterized by ensemble covariances rather than fixed offsets.
We extend Eq.~\eqref{eq:R_statistic_equations} in the following section to account for such glitch-related stochasticity.

\subsection{Biases from glitch backgrounds}\label{Sec:Biases from glitch backgrounds}

From Eq.~\eqref{eq:sum_of_MLEs_glitches_noise}, the combined noise- and glitch-induced contributions to the MLE can be written as
\begin{equation}
    \label{eq:glitch_error_eqn}
    \Delta\theta^{i}
    = \left(\Gamma^{-1}\right)^{ij}\left(\partial_j h|n + g\right).
\end{equation}

The fact that the noise and glitch backgrounds are stochastic processes implies that Eq.~\eqref{eq:glitch_error_eqn} is \emph{also} stochastic. 
We can calculate the bias of the MLE by taking the expectation of Eq.~\eqref{eq:glitch_error_eqn}.
For interpretability, we also make it relative to the noise-induced uncertainty to obtain the statistic $\beta^{i}$ which has the form
\begin{equation}\label{eq:beta_equation}
\beta^{i} = \frac{\mathbb{E}[\Delta\theta^{i}_{\text{glitches}}]}{\Delta\theta^{i}_{\text{stat}}}\,,
\end{equation}
and so we express the bias in terms of the number of standard deviations from the true parameters.
In effect, Eq.\eqref{eq:beta_equation} repurposes Eq.\eqref{eq:R_statistic_equations} by swapping the waveform mismodeling error $h_{\text{e}} - h_{\text{m}}$ with a glitch background, and allowing the MLE fluctuations to be stochastic rather than deterministic.

When the data consists only of a GW signal and noise, the parameter covariance reduces to Eq.~\eqref{eq:covariance_noise_stat}.
Including glitches in the data causes the covariance to increase, for which we can derive a corrected form of  Eq.~\eqref{eq:covariance_noise_stat}.
Following Ref.~\cite{Burke:2025bun} and using the discrete inner product of Eq.~\eqref{eq:discrete_inner_product}, we derive an analytical expression for the MLE covariance from Eq.~\eqref{eq:glitch_error_eqn}.
The result is given by 
\begin{widetext}
\begin{align}\label{eq:covariance_glitch_biases_main_text}
\text{Cov}[\widehat{\Delta\theta^{i}},\widehat{\Delta\theta^{j}}] & = \mathbb{E}[\widehat{\Delta\theta^{i}}\widehat{\Delta\theta^{j}}] - \mathbb{E}[\widehat{\Delta\theta^{i}}]\mathbb{E}[\widehat{\Delta\theta^{j}}] \nonumber \\
& \approx (\Gamma^{-1})^{ik}(\Gamma^{-1})^{jl}\{\mathbb{E}[((\partial_kh|g + n)(n + g|\partial_{l}h)] - \mathbb{E}[(\partial_k h|g)]\mathbb{E}[(\partial_lh|g)]\}\nonumber \\
&\approx  2(\Gamma^{-1})^{ik}(\Gamma^{-1})^{jl}\bigg\{\mathrm{Re}\left((\partial_k \boldsymbol{\tilde{h}})^T\boldsymbol{\Sigma}^{-1}\boldsymbol{C}_{g}\boldsymbol{\Sigma}^{-1}\partial_l \boldsymbol{\tilde{h}}\right) + \mathrm{Re}\left((\partial_k \boldsymbol{\tilde{h}})^\dagger\boldsymbol{\Sigma}^{-1}\{\boldsymbol{\Sigma}_{g} + \boldsymbol{\Sigma}\}\boldsymbol{\Sigma}^{-1}\partial_l \boldsymbol{\tilde{h}}\right) \\
& \qquad \qquad \qquad \qquad \qquad - \bigg[\text{Re}\bigg((\partial_k \boldsymbol{\tilde{h}})^T \boldsymbol{\Sigma}^{-1} \boldsymbol{\tilde{\bar{g}}} \boldsymbol{\tilde{\bar{g}}}^T \boldsymbol{\Sigma}^{-1} \partial_l \boldsymbol{\tilde{h}}\bigg) + \text{Re}\bigg((\partial_k \boldsymbol{\tilde{h}})^\dagger \boldsymbol{\Sigma}^{-1} \boldsymbol{\tilde{\bar{g}}} \boldsymbol{\tilde{\bar{g}}}^\dagger \boldsymbol{\Sigma}^{-1} \partial_l \boldsymbol{\tilde{h}}\bigg) \bigg]\bigg\}\, .
\end{align}
\end{widetext}
Here $\mathbf{C}_{g} = \mathbb{E}[\boldsymbol{\tilde{g}}\boldsymbol{\tilde{g}}^{T}]$ and $\mathbf{\Sigma}_{g} = \mathbb{E}[\boldsymbol{\tilde{g}}\boldsymbol{\tilde{g}}^{\dagger}]$ denote the pseudo-covariance and covariance matrices of the glitch background respectively. 
We also define $\mathbb{E}[\tilde{\boldsymbol{g}}] = \boldsymbol{\tilde{\bar{g}}}$. 
In evaluating Eq.~\eqref{eq:covariance_glitch_biases_main_text}, we use the fact that the pseudo-covariance matrix of the noise vanishes over positive frequencies, that $\boldsymbol{\Sigma}$ is real and diagonal (ensuring Hermitian symmetry), and that $\mathbb{E}[n] = 0$. 
The noise and glitch processes are assumed to be statistically independent. 
In the absence of glitches, where $\mathbf{C}_{g} = \mathbf{\Sigma}_{g} = \boldsymbol{\tilde{\bar{g}}} = 0$, Eq.~\eqref{eq:covariance_glitch_biases_main_text} reduces to the standard noise-induced covariance of Eq.~\eqref{eq:covariance_noise_stat}.

In Eq.~\eqref{eq:covariance_glitch_biases_main_text}, the first two terms represent statistical fluctuations in the parameters around their mean values arising from both glitches and noise. 
The latter two terms correspond to bias correction introduced by non-zero glitch populations. 
As seen in Tab.~\ref{tab:glitch_impact_summary} of Sec.~\ref{sec:results}, numerical tests with weak low-SNR glitch backgrounds based on the LPF catalog indicate that this contribution is negligible. 
In contrast, strong glitch backgrounds resulted in both a constant bias offset (set by the mean glitch amplitude) and an increase in the parameter covariance.

The fact that glitches are localized in time implies that they are non-local in frequency, resulting in dense covariance matrices $\mathbf{C}_{g}$ and $\mathbf{\Sigma}_{g}$ which makes Eq.~\eqref{eq:covariance_glitch_biases_main_text} computationally expensive for very long time-series.
Consequently, applying Eq.~\eqref{eq:covariance_glitch_biases_main_text} in this work proves to be impractical.
We instead employ a simpler numerical approach discussed in Ref.~\cite{Burke:2025bun} using Monte-Carlo methods.
Specifically, we generate many realizations of glitch and noise backgrounds, estimate the corresponding MLEs, and hence estimate the parameter covariance defined by Eq.~\eqref{eq:glitch_error_eqn} using the distribution of MLEs.

We therefore define the principal statistic $\Upsilon_{ij}$, dubbed the \emph{scatter-to-width} ratio, to estimate the glitch-induced multiplicative change in parameter covariance:
\begin{equation}\label{eq:upsilon_matrix}
\Upsilon^{ij} = \frac{\text{Cov}[\widehat{\Delta\theta^{i}},\widehat{\Delta\theta^{j}}]}{(\Gamma^{-1})^{ij}}\,.
\end{equation}
The diagonal elements quantify the inflation of a given parameter's uncertainties due to glitch backgrounds.
When $\Upsilon^{ii}\simeq1$, there is negligible change in the covariance, implying that parameter fluctuations remain consistent with the underlying noise model. 
This corresponds to the limit $\boldsymbol{\Sigma}_{g}\approx\boldsymbol{C}_{g}\approx \boldsymbol{\tilde{\bar{g}}} \approx 0$ in Eq.~\eqref{eq:covariance_glitch_biases_main_text}. 
Conversely, $\Upsilon^{ii}>1$ indicates that parameter uncertainties are being inflated by some form of model misspecification.

The statistics $\beta^{i}$ and $\Upsilon^{ij}$ defined by Eqs.\eqref{eq:beta_equation} and \eqref{eq:upsilon_matrix}
serve as measures of location and spread for MLEs arising from stochastic processes such as glitches.
For the rest of this work, $\beta^{i}$ and $\Upsilon^{ij}$ serve as our primary diagnostics for quantifying glitch-induced inference bias and uncertainty.
For the interested reader, more theoretical discussion of Eq.~\eqref{eq:upsilon_matrix} can be found in Sec.~IV~D of Ref.~\cite{Burke:2025bun}.

\section{Modeling EMRIs and Transient Noise}\label{sec:modelling}

\subsection{Waveforms and glitches}
\label{sec:waveforms_glitches}

We model EMRIs using the Kerr Eccentric Equatorial waveform family from \texttt{FastEMRIWaveforms}~\cite{Chua:2020stf,Katz:2021yft,michael_l_katz_2023_8190418,Chua:2018woh,Chua:2017ujo,Speri:2023jte,chapmanbird2025fastframedraggingefficientwaveforms}. 
This model describes the gravitational radiation from a non-spinning compact object of mass~$\mu$ inspiralling on an eccentric, equatorial orbit around a Kerr black hole of mass~$M$ and spin~$a$. 
The intrinsic parameters are $\{M,\mu,a,e_0,p_0,\Phi_{\phi_0},\Phi_{r_0}\}$, where $e_0$ and $p_0$ denote the initial eccentricity and semi-latus rectum, and $\Phi_{\phi_0},\Phi_{r_0}$ are initial orbital phases~\cite{Lynch_2025}. 
The extrinsic parameters are $\{D_{\mathrm{L}},\theta_S,\phi_S,\theta_K,\phi_K\}$, specifying luminosity distance, sky location, and spin orientation respectively. 
We fix $(x_I)_0=\cos\iota_0=1$ and $\Phi_{\theta_0}=0$.

In this work, we look at three fiducial EMRIs: a canonical (prograde), retrograde, and strong-field system~\cite{chapmanbird2025fastframedraggingefficientwaveforms}, whose parameters are listed in Table~\ref{tab:3_EMRI_params}. 
They span complementary regions of EMRI parameter space, capturing different mass ratios, orbital frequencies, and spin orientations relevant to LISA. 
Each source is observed for $T_{\mathrm{obs}}=2~\mathrm{yr}$ with luminosity distance $D_{\mathrm{L}}$ adjusted to yield an optimal SNR of~80. 
The initial semi-latus rectum $p_0$ is tuned so that the plunge occurs just after the observation window. 
Sampling cadences are $\Delta t=5~\mathrm{s}$ for the prograde and strong-field systems, and $2~\mathrm{s}$ for the higher-frequency retrograde system to capture the high frequency content of the signal. 

\begin{table}[t]
    \centering
    \renewcommand{\arraystretch}{1.05}
    \setlength{\tabcolsep}{6pt}
    \begin{tabular}{l c c c c c}\toprule
        % \hline\hline
        EMRI & $M\,[M_{\odot}]$ & $a$ & $p_0$ & $e_0$ & $D_\mathrm{L}\,[\rm{Gpc}]$ \\\midrule
        
        Canonical & $10^6$ & $0.998$ & $7.728$ & $0.73$ & $2.20$\\
        Strong-field & $10^7$ & $0.998$ & $2.12$ & $0.43$ & $3.59$\\
        Retrograde & $10^5$ & $-0.5$ & $26.19$ & $0.80$ & $1.08$\\ \bottomrule
        % \hline\hline
    \end{tabular}
    \caption{
    Fiducial EMRIs used in this study.  
    Each system has secondary mass $\mu = 10\,M_\odot$, optimal SNR $\rho_{\mathrm{opt}} = 80$, and plunges just beyond the observing window $T_{\mathrm{obs}} = 2~\mathrm{yr}$.  
    Extrinsic parameters are fixed to $(\theta_S,\phi_S,\theta_K,\phi_K,\Phi_{\phi_0},\Phi_{r_0})=(0.8,2.2,1.6,1.2,2.0,3.0)$.    \label{tab:3_EMRI_params} }
\end{table}

We adopt a static, equal-arm LISA configuration with constant $2.5\times 10^6~\mathrm{km}$ arm lengths, allowing construction of second-generation time-delay interferometry (TDI) observables $\{X,Y,Z\}$.
We use the corresponding orthogonal channels $\{A,E,T\}$~\cite{Katz_2022} which have the form
\begin{subequations}
\begin{align}\label{eq:transform_XYZ_to_AET}
A &= (Z-X)/\sqrt{2}, \\
E &= (X-2Y+Z)/\sqrt{6}, \\
T &= (X+Y+Z)/\sqrt{3}.
\end{align}
\end{subequations}
For our analysis, we disregard the $T$ channel due to its insensitivity to EMRIs (though not to glitches) and assume identical, uncorrelated noise PSDs $S_n^{A}=S_n^{E}$.

TDI propagation is implemented with the GPU-accelerated \texttt{fastlisaresponse}~\cite{Katz_2022}.  
Stationary instrumental noise follows the \texttt{SciRDv1} sensitivity curve~\cite{scirdv1}, with the unresolved galactic binary background being simulated using \texttt{lisatools}~\cite{michael_katz_2024_10930980}.

We model each glitch's base interferometer measurements with the integrated shapelet model~\cite{Berg__2019,PhysRevD.105.042002}. 
Glitch $i$ is parameterized by vector $\boldsymbol{\lambda}^{(i)}=(\alpha,\beta,\tau)$ whose elements denote the amplitude~$\alpha$, decay time~$\beta$, and onset~$\tau$ respectively. 
The corresponding velocity perturbation to the test-mass within the Moving Optimal Sub Assembly (MOSA), (see Fig.\ref{app_fig:instrument_config} in Appendix~\ref{app_sec:noise_covariance_glitches_interferometry}) is
\begin{equation}
\label{integrated_shapelet}
     v_g(t;\alpha,\beta,\tau)
     =
     2 \alpha \beta \left[1 - \left(1 + \frac{\delta t}{\beta}\right) e^{-\delta t / \beta}\right]H(t-\tau),
\end{equation}
where $\delta t = t - \tau$ is the time since glitch onset, and $H(t-\tau)$ is the Heaviside function, and multiple independent glitches are summed as
$g(t;\boldsymbol{\Lambda})=\sum_i g(t;\boldsymbol{\lambda}^{(i)})$,  
with parameters $\boldsymbol{\lambda}^{(i)}$ drawn from LPF-informed distributions.

Glitches are injected into LISA's test-mass interferometer using \texttt{lisaglitch}~\cite{bayle_2022_6452904} and propagated through TDI with the same response model as for EMRIs (see Appendix~\ref{app_sec:noise_covariance_glitches_interferometry}). 
Their brightness is quantified through the network optimal SNR, 
\begin{align}
\rho^2_{A,E} =
4\int_{0}^{\infty}
\!\left(
\frac{|\tilde g_A(f)|^2}{S^A_n(f)}+
\frac{|\tilde g_E(f)|^2}{S^E_n(f)}
\right)\!{\rm d}f,
\label{eq:detector_SNR_glitches}
\end{align}
computed using the \texttt{SciRDv1} noise curve.

\begin{figure*}
    \centering
    \includegraphics[width=\linewidth]{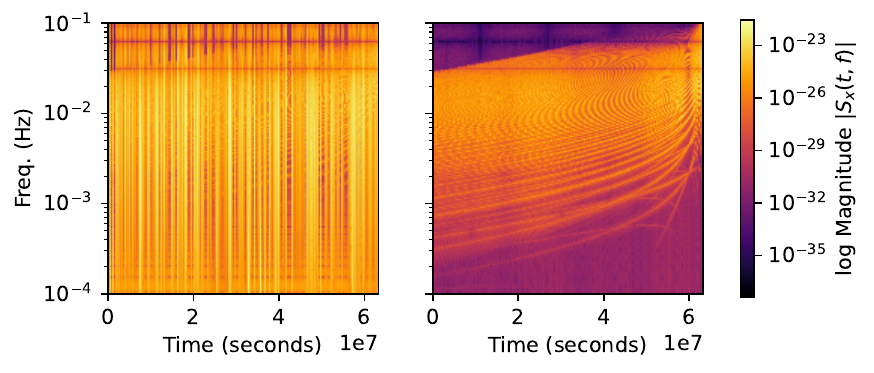}
    \caption{A Short-Time Fourier Transform spectrogram of the TDI response of the prograde EMRI buried in an unmitigated glitch background (left), alongside the pure EMRI (right). We used a Tukey window of length $32,768$ samples and an overlap of $75\%$. This was to ensure a relatively high frequency resolution, making the EMRI modes (seen as arching diagonal lines) easy to distinguish without making the glitches (seen as sharp vertical lines) poorly resolved in time.}
    \label{fig:glitchy_EMRI_spectrogram}
\end{figure*}

In Figure~\ref{fig:glitchy_EMRI_spectrogram}, we visualize how glitches obscure EMRIs in the time-frequency domain.
The diagonal tracks reveal the time and frequency evolution of the individual EMRI harmonics, while the sharp vertical lines indicate glitch placement over the 2 year long observation.

\begin{figure}
    \centering
    \includegraphics[width=\linewidth]{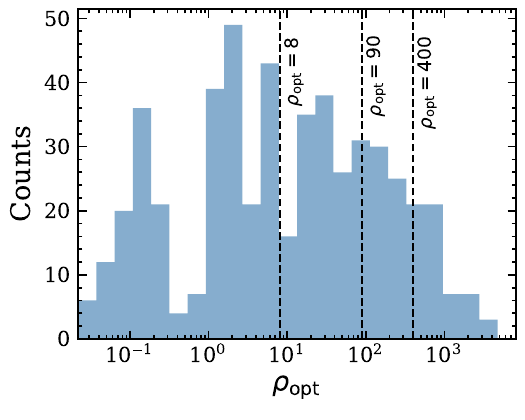}
    \caption{Distribution of network optimal SNRs for individual LISA-injected glitches from the LPF catalog, computed using the second-generation \texttt{SciRDv1} noise PSD. Vertical dashed lines indicate our adopted glitch mitigation thresholds at $\rho_{\mathrm{A,E}} = \{8,90,400\}$, corresponding approximately to the 50th, 75th, and 90th percentiles of the glitch $\rho_{\mathrm{A,E}}$ distribution.}
    \label{fig:glitch_catalog}
\end{figure}

Figure~\ref{fig:glitch_catalog} shows the SNR distribution of LISA-injected glitches from the LPF catalog: some $50\%$ have up to moderate amplitudes ($\rho\!\lesssim\!10$), while a minority exceed $\rho\gtrsim10^3$.
Very loud glitches such as these would likely be mitigated or masked in practice.
In the next section, we use the annotated percentiles in Figure~\ref{fig:glitch_catalog} to construct ``mitigated'' glitch backgrounds to be used in our estimation of inference bias and covariance.

\subsection{Mitigated glitch backgrounds}
\label{Glitch backgrounds and mitigation}

We model glitch backgrounds as streams of integrated shapelet glitches whose $(\alpha,\beta)$ parameters are resampled with replacement from the LPF glitch catalog \cite{lpf_glitches}.
The injection times $\tau$ are drawn uniformly across all six test-mass interferometers.  
Using the $\simeq1$~day$^{-1}$ glitch rate, we treat the glitch count in each background as a homogeneous Poisson process averaging at $730$ glitches per two years.  
We exclude glitches with individual $\rho_{\rm opt}>10^4$, corresponding to extremely bright glitches that would be easy to identify.  

To mimic the output of a glitch mitigation algorithm, we generate a stream of glitches ``mitigated'' by omitting every glitch whose individual $\rho_{\rm opt}$ exceeds some threshold e.g. $\{8,90,400\}$.  
This assumes perfect identification and subtraction of glitches.
Although optimistic, this approach  does provide an upper bound on the effectiveness of mitigation.  
A more realistic treatment would involve joint inference on overlapping glitches or FM estimates of individual glitch MLEs in the presence of an EMRI signal, which we defer to future work.

For this study, we construct $N=128$ independent glitch backgrounds, each mitigated to single-glitch SNRs $\rho_{\rm opt}>\{8,90,400\}$, corresponding approximately to the 50th, 75th, and 90th percentiles of the SNR distribution in Fig.~\ref{fig:glitch_catalog}.  
These represent decreasingly strict mitigation levels, allowing us to quantify the dependence of EMRI PE on mitigation efficiency.

Each glitch background is propagated through TDI following Appendix~\ref{app_sec:noise_covariance_glitches_interferometry}, and its total network SNR is computed using Eq.~\eqref{eq:detector_SNR_glitches}.  
We find that unmitigated backgrounds have network SNRs clustered around $\rho_{\text{net}}\!\sim\!2\times10^{4}$.
After mitigating glitches in excess of $\rho_{\rm opt}>\{8,90,400\}$, the average network SNRs shift to $\rho_{\text{net}}\!\sim\{ 10^{2},  10^{3},\!4\times10^{3}\}$, respectively.

\section{Results}\label{sec:results}

We demonstrate and present the impact of glitch backgrounds on EMRI inference by estimating the parameter bias and uncertainty induced by glitch backgrounds mitigated to varying degrees.
For parameter-wise comparison, we normalize the bias and uncertainty by the Fisher-derived, noise-induced uncertainty $\Delta\theta_{\text{stat}}^i$.
This also allows comparisons between glitch-induced biases and the statistical limits imposed by Gaussian noise.

Across $128$ glitch realizations, we compute the normalized ensemble-averaged bias and uncertainty inflation using \begin{align} \beta^i &= \frac{\mathbb{E}[\widehat{\Delta\theta^i}]}{\Delta\theta_{\text{stat}}^i}, & \sigma^i &= \sqrt{\Upsilon^{ii}} = \frac{\sqrt{\mathrm{Var}[\widehat{\Delta\theta^i}]}}{\Delta\theta_{\text{stat}}^i}. \label{eq:beta_and_1sigma_mismodelling}\end{align} 
The latter term in Equation \eqref{eq:beta_and_1sigma_mismodelling} represents the square-rooted diagonal elements of the normalized mismodeled parameter covariances due to glitch backgrounds.
We do not report the off-diagonal elements in \eqref{eq:upsilon_matrix} since they do not impact our main conclusions.

Using the metrics defined in Eq.~\eqref{eq:beta_and_1sigma_mismodelling}, we quantify how glitches degrade inference for the canonical (prograde), retrograde, and strong-field EMRIs under four cases of mitigation corresponding to decreasingly strict single-glitch SNR thresholds of $\{8, 90, 400, \infty\}$.

Table~\ref{tab:glitch_impact_summary} summarizes the key statistics across all three fiducial EMRIs. 
All systems exhibit similar qualitative trends, with quantitative variations reflecting their distinct orbital dynamics and spectral support.
Compared to the prograde EMRI, the strong-field one exhibits larger biases and higher uncertainty inflation  due to its stronger high-frequency content.
The retrograde EMRI also shows higher biases and a notably wide range of uncertainties linked to its broader TDI response overlap.
In unmitigated scenarios, the largest biases exceed the $1\sigma$ level, directly compromising high-precision tests of GR.

\begin{table*}[t]
\centering
\caption{Summary of the largest absolute glitch-induced biases $\max|\beta^i|$ and the ranges of uncertainty inflation $(\min\sigma^i,\max\sigma^i)$ for the three fiducial EMRIs at $\rho_{\mathrm{opt}}=80$. Biases are expressed in units of noise-only $1\sigma$ standard deviations.}
\setlength{\tabcolsep}{6pt} % Increased horizontal spacing from 4pt to 10pt (default is 6pt)
\renewcommand{\arraystretch}{1.3} % Increased vertical spacing by 30%
\begin{tabular}{p{0.1\linewidth}p{0.1\linewidth}p{0.1\linewidth}|cc|cc|cc}
& & & \multicolumn{2}{c}{Canonical (Prograde)} & \multicolumn{2}{c}{Retrograde} & \multicolumn{2}{c}{Strong-field} \\
\cline{4-9}
Mitigation Threshold & Avg. mitigated ($\%$) & Avg. background SNR & $\max|\beta^i|$ & Range of $\sigma^i$ & $\max|\beta^i|$ & Range of $\sigma^i$& $\max|\beta^i|$ & Range of $\sigma^i$ \\
\hline
SNR $>8$ & 50 & $\sim10^2$ & $0.007$ & $\approx1$ & $0.1$ & $(1.0,3.0)$ & $0.01$ & $\approx1$ \\
SNR $>90$ & 25 & $\sim10^3$ & $0.04$ & $(1.05,1.24)$ & $0.6$ & $(1.0,13.0)$ & $0.07$ & $(1.23,1.37)$ \\
SNR $>400$ & 10 & $\sim4\times10^3$ & $0.3$ & $(1.9,3.0)$ & $0.4$ & $(1.6,44.0)$ & $0.7$ & $(3.0,5.0)$ \\
No mitigation & 0 & $\sim2\times10^4$ & $1$ & $(3.6,8.5)$ & $5$ & $(4.1,73.0)$ & $3$ & $(8.7,12.6)$ \\
\end{tabular}
\label{tab:glitch_impact_summary}
\end{table*}

For all three EMRIs, the maximum absolute biases never exceeded $0.7\sigma$ except in the case of unmitigated data.
This establishes that mitigating glitches of SNR $>400$ only results in moderately biased inference.
More stringent glitch mitigation at SNR thresholds $>90$ or $>8$ is preferable for minimizing the biases.
In all cases, a lack of glitch mitigation resulted in biases exceeding $\pm1\sigma$ meaning that we cannot reasonably perform EMRI inference in this scenario.
However, we do remark that search pipelines whose primary focus is to identify a subregion close to the true parameters may be less affected by lax (if any) glitch mitigation.

For the prograde and strong-field EMRIs, uncertainties remain near unity when mitigating above SNR $90$, while unmitigated data exhibit uncertainties that are several times larger than the noise-only case.
This means that parameter fluctuations are dominated by the noise process and not the treated glitch process.
The retrograde EMRI, having stronger overlap between its harmonics and TDI response, is much more sensitive to glitches and exhibits a much wider range of uncertainties at all mitigation levels.
These results suggest that more stringent glitch mitigation e.g. as low as SNR $>8$ is necessary to avoid imprecise PE. 

\begin{figure} 
\centering \includegraphics[width=\linewidth]{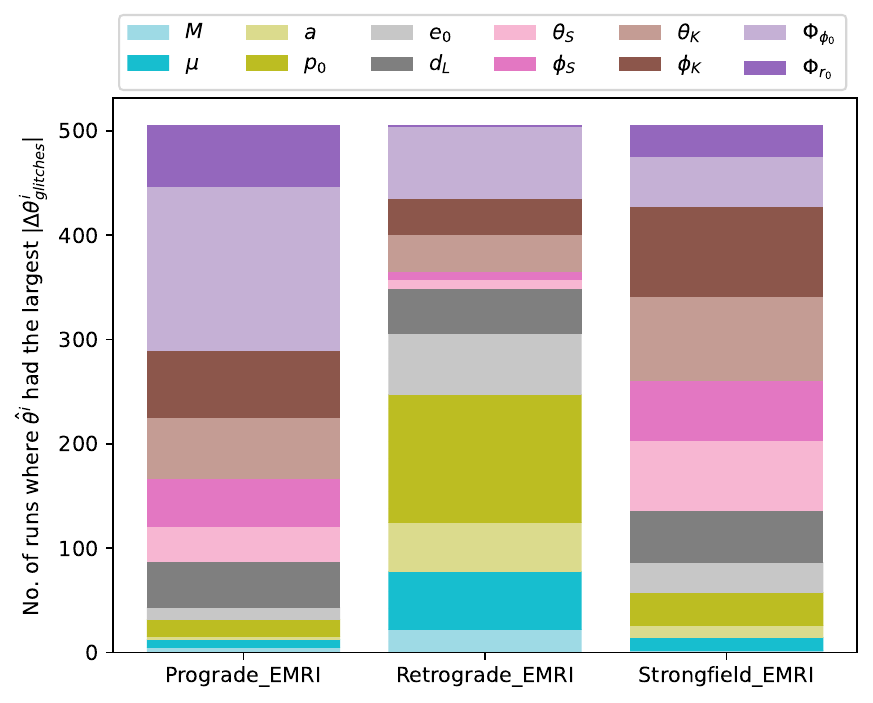} 
\caption{Each EMRI's count of runs in which a given parameter had the largest absolute normalized statistical fluctuation i.e. $\arg \max |\Delta\theta_{\text{glitches}}^i/\Delta\theta_{\text{stat}}^i|$. The dotted colors represent intrinsic parameters. By proportion, each bar is the same as $\mathbb{P}\left(\arg\max_i \frac{|\Delta\theta_{\text{glitches}}^i|}{\Delta\theta_{\text{stat}}^i}=i\middle|g\right)$. We found little to no variation in results across glitch mitigation levels, so the counts are combined across all mitigation levels.
For the prograde and strong-field EMRIs, extrinsic parameters dominate, consistent with their weaker Fisher constraints and susceptibility to phase perturbations.}
\label{fig:EMRI_bias_piechart} 
\end{figure}

For each fiducial EMRI, we also wished to understand which inferred parameters in a given run were most likely to have the largest glitch-induced fluctuation.
For this we computed $\arg \max |\Delta\theta_{\text{glitches}}^i/\Delta\theta_{\text{stat}}^i|$ for each run at each mitigation level, and present the combined counts in Fig.~\ref{fig:EMRI_bias_piechart}.
For the prograde and strong-field EMRIs, extrinsic parameters such as the sky position $(\phi_K)$ and phase angles $(\Phi_{\phi_0}, \Phi_{r_0})$ were overwhelmingly likely to have the largest statistical fluctuations. 
In the retrograde case, we find instead that intrinsic parameters such as the initial semi-latus rectum ($p_0$) dominate.
It remains uncertain to us why this was the case.
We hypothesize that the retrograde EMRI's poorer precisions on the instrinsic parameters render it more prone to glitch-related parameter flucutations.

\section{Discussions and Conclusions}
\label{sec:conclusions}

The results presented in this work demonstrate that transient, non-Gaussian “glitches” can measurably bias EMRI PE if left unmitigated, but that these effects are largely controllable with realistic levels of glitch removal.
Building on the FM framework developed in Sec.~\ref{sec:fisher_matrix}, we quantified how streams of LPF-like glitches degrade the unbiasedness and precision of EMRI inference under different mitigation strategies.

For LISA-based EMRI signals with $\rho_{\mathrm{opt}}\!\gtrsim\!80$, streams of LPF-like glitches mitigated above SNRs $\sim90$ introduce negligible-to-moderate biases ($|\beta|<0.6\sigma$) and usually minor inflation of parameter uncertainties.
This means that even when the cumulative glitch SNR reaches $\sim\!10^{3}$, reasonably unbiased and precise inference is possible.
In contrast, lax mitigation strategies that retain glitches of SNRs $\lesssim\!400$ produce systematic biases approaching $1\sigma$ and inflate uncertainties by factors of $O(1)$.
This highlights the importance of effective transient identification and subtraction in LISA data analysis pipelines.
Overall, these results indicate that EMRI PE can remain unbiased and precise without requiring exhaustive amounts of glitch mitigation.

We found that in the prograde and strong-field EMRIs, the largest glitch-induced fluctuations in a given run were most likely to be in extrinsic parameters such as $\Phi_{\phi_0}$, $\Phi_{r_0}$, and $\phi_K$ which govern the source’s sky orientation and phase evolution.
Such distortions could compromise sky localization and distance estimation, thereby impacting multi-messenger and cosmological analyses that rely on joint electromagnetic and GW identification of EMRI host galaxies~\cite{derdzinski2023multimessengerastronomyblackholes}.
We also found that the retrograde EMRI's largest glitch-induced fluctuations were more likely to be from intrinsic parameters such as $a$, $\mu$, and $p_0$, which could challenge EMRI-based tests of GR.

It is important to acknowledge the simplifying assumptions underlying this study.
The FM formalism provides a computationally efficient approximation valid only in the high-SNR, near-linear regime, but may underestimate correlations or non-Gaussianity in the true posterior.
Realistic LISA observations will also contain quieter EMRIs, additional and overlapping sources, and time-dependent instrumental noise.
We note that LISA could experience other kinds of glitches such as sine-Gaussians whose morphology and statistics could further impact the parameter biases and covariances beyond what is reported here.

We also reiterate that our study did not account for the $T$-channel contribution.
Given the channel's insensitivity to EMRIs but not glitches, we conjecture that including the $T$-channel would marginally increase the estimated biases.
One other consideration is our assumption of exact glitch mitigation.
In practice, glitch mitigation algorithms will leave non-zero residuals in the data which would further shift the biases.
Investigating the impact of these residual-induced biases would be a fitting avenue for future research.
% In the case of matching background SNRs, the equivalence between perfect mitigation of a subset (as done in our study) and imperfect mitigation of the full glitch population remains unestablished.
% Investigating the impact of these residuals represents a primary objective for future research.

Within our controlled assumptions, our findings suggest that moderate mitigation strategies targeting only the strongest transients may suffice to preserve near-optimal inference accuracy and precision while minimizing data loss.
This is convenient as glitch mitigation can be technically challenging, requiring precise modeling of transient morphology and careful separation from overlapping astrophysical signals.
Extending this framework to existing glitch mitigation algorithms, a larger and broader population of EMRIs, and non-stationary or correlated noise models will be essential for assessing the robustness of EMRI science in the LISA era.
We reiterate that the validity of our approach is limited to high-SNR EMRIs, and so the inference bias and uncertainty of quiet EMRIs remains unexplored.
Ultimately, future EMRI-related glitch studies will help ensure that strong-field tests of GR and cosmological measurements with LISA remain resilient to realistic instrumental and astrophysical noise environments.

\begin{acknowledgments}
We thank the LISA Simulation Expert Group, and in particular the authors of LISA Glitch: Jean-Baptiste Bayle, Eleonora Castelli, and Natalia Korsakova. 
We thank again Jean-Baptiste Bayle for his assistance in propagating glitches through \texttt{fastlisaresponse}.
We thank Ruiting Mao for initial discussions and exploration on the glitch model.  
This work was performed on the OzSTAR national facility at Swinburne University of Technology.
The OzSTAR program receives funding in part from the Astronomy National Collaborative Research Infrastructure Strategy (NCRIS) allocation provided by the Australian Government, and from the Victorian Higher Education State Investment Fund (VHESIF) provided by the Victorian Government. 

A.B., M.C.E., and A.V. kindly acknowledge support by the Marsden Grant No. MFP-UOA2131 from New Zealand Government funding, administered by the Royal Society Te Apārangi.  

M.C.E. and O.B. acknowledge the University of Auckland's Faculty of Science Research Development Fund No. 3726778.
O. Burke also acknowledges financial support from the Grant UKRI972 awarded via the UK Space Agency.

All work was done in \texttt{Python} using \texttt{NumPy}~\cite{harris2020array}, \texttt{CuPy}~\cite{cupy_learningsys2017}, \texttt{SciPy}~\cite{2020SciPy-NMeth}, \texttt{Corner}~\cite{corner}, \texttt{scikit-learn}~\cite{scikit-learn}, and \texttt{Matplotlib}~\cite{Hunter:2007}.

We provide the plotting scripts for this paper at \hyperlink{https://github.com/nz-gravity/emri_glitch_paper_datasets}{the following GitHub repository}, while data generation and analysis code can be provided upon reasonable request.

\end{acknowledgments}

\appendix

\section{TDI response of test-mass interferometer glitches}\label{app_sec:noise_covariance_glitches_interferometry}
\begin{figure}
\centering
\includegraphics[width=0.45\textwidth]{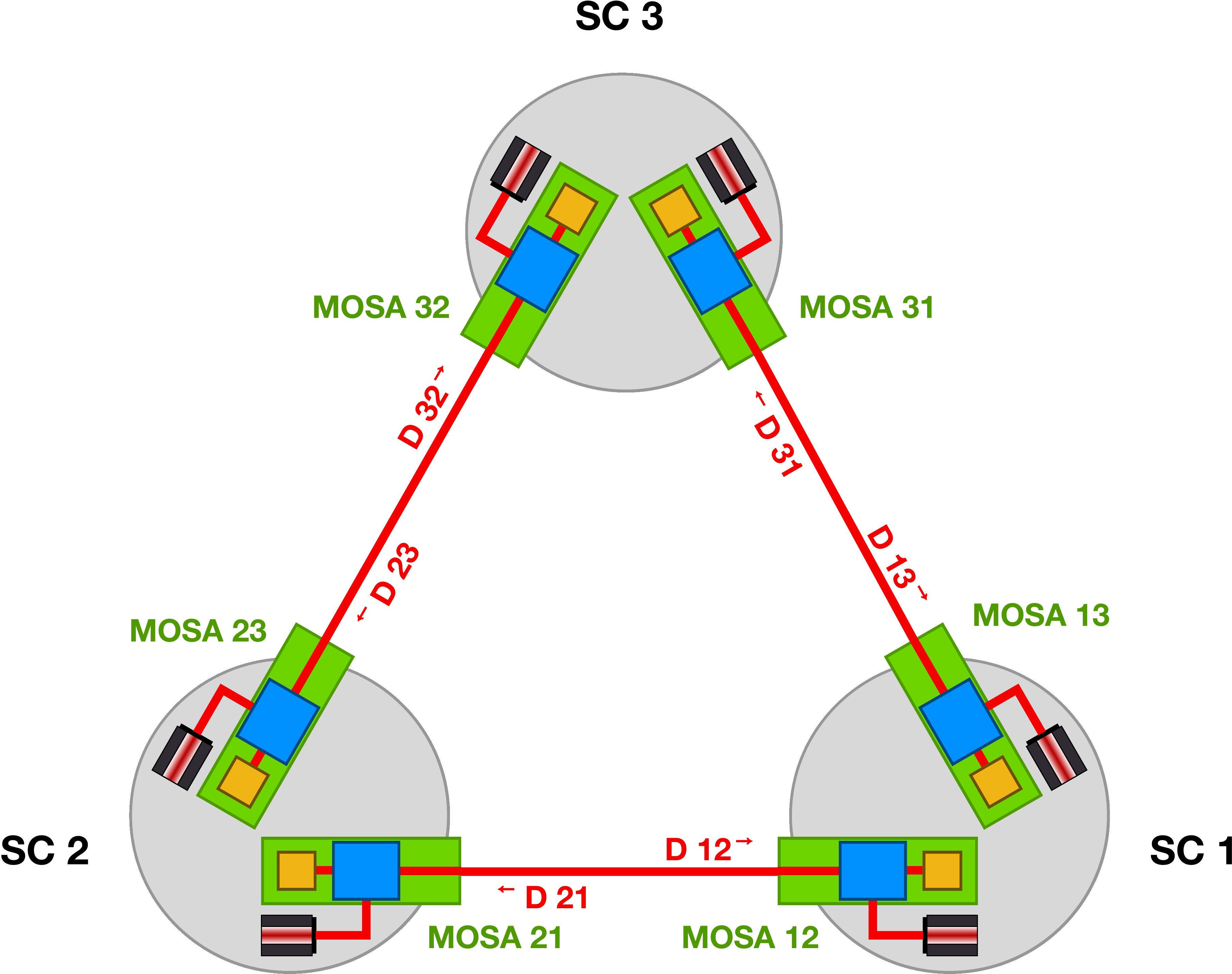}
\caption{Schematic diagram taken from~\cite{PhysRevD.107.083019} of the LISA constellation showing the three spacecraft (S/C 1, S/C 2, S/C 3) and their respective moving sub-optical assemblies (MOSA). Each MOSA is labeled with indices $(i,j)$ where $i$ denotes the host spacecraft and $j$ indicates the spacecraft from which it receives laser light. Red arrows indicate the inter-spacecraft links with associated light-travel distances $D_{ij}$.}
\label{app_fig:instrument_config}
\end{figure}

We calculate the TDI response for glitch signals injected into the test-mass interferometers, formulated to interface directly with \texttt{fastlisaresponse} for efficient TDI computation.
The LISA constellation comprises three spacecraft, each housing two lasers and two optical benches~\cite{colpi2024lisadefinitionstudyreport}.
We label each spacecraft with an index from $\{1,2,3\}$ and denote each moving sub-optical assembly (MOSA) by an ordered pair $(i,j)$, where $i$ identifies the host spacecraft and $j$ specifies the spacecraft providing the incoming laser beam (see Fig.~\ref{app_fig:instrument_config}).
The fundamental observables in LISA are the phase measurements $\text{isi}_{ij}$, $\text{tmi}_{ij}$, and $\text{rfi}_{ij}$ recorded by the inter-spacecraft, test-mass, and reference interferometers, respectively.
Time-delay interferometry (TDI) synthesizes these measurements by combining appropriately time-delayed signals from multiple locations throughout the instrument.

We define the delay operator $\textbf{D}_{ij}$ which delays a specific measurement $x(t)$ by the light travel time $L_{ij}/c$ between spacecrafts $i$ and $j$ such that for some measurement $x(t)$
\begin{equation}
    \textbf{D}_{ij}x(t)=x\left(t-\frac{L_{ij}}{c}\right).
\end{equation}
For our analysis we will consider a static constellation, implying that the arm-length between individual S/Cs $L_{ij} = L$ with non-evolving arm-lengths $\dot{L} = 0$. This means we can write $\textbf{D}_{ij} = \textbf{D}$ with 
\begin{equation}
    \textbf{D}x(t)=x\left(t-\frac{L}{c}\right).
\end{equation}
with multiple (nested) delay operators obeying the identity $\mathbf{D}_{i_1i_2}\mathbf{D}_{i_2i_3}\dots\mathbf{D}_{i_{k-1}i_k} = \mathbf{D}_{i_1i_2\dots i_k} = \mathbf{D}^{k-1}$, with e.g operation
\begin{align}
    \textbf{D}_{ijk}x(t)=\textbf{D}_{ij}\textbf{D}_{jk}x(t) &= x\left(t-2\frac{L}{c}\right). \\
    &= \textbf{D}^2 x(t). 
\end{align}

For each the two optical benches per MOSA, there are associated intermediary variables $\xi_{ij}$ and $\eta_{ij}$ constructed to mitigate spacecraft jitter and to reduce the 6 laser links to three independent effective links. The $\xi_{ij}$ combinations have the form
\begin{equation}
    \xi_{12}
    =
    \text{isi}_{12}
    +
    \frac{\text{rfi}_{12}-\text{tmi}_{12}}{2}
    +
    \frac{\textbf{D}(\text{rfi}_{21}-\text{tmi}_{21})}{2},
\end{equation}
whose subsequent index combinations $(ij)$ are obtained by rotation and reflection of the indices.
The $\eta_{ij}$ combinations have the form
\begin{equation}
    \eta_{12}
    =
    \xi_{12}
    +
    \frac{\textbf{D}(\text{rfi}_{21}-\text{rfi}_{23})}{2},
\end{equation}
and
\begin{equation}
    \eta_{13}
    =
    \xi_{13}
    +
    \frac{\text{rfi}_{12}-\text{rfi}_{13}}{2},
\end{equation}
whose subsequent index combinations are obtained by cyclic index permutation.

We can then express the first generation TDI variables $\{X_1,Y_1,Z_1\}$ for a static LISA configuration
\begin{align}
    X_1
    \approx
    (1-\textbf{D}^2)[\eta_{13}+\textbf{D}\eta_{31} - \eta_{12} - \textbf{D}\eta_{21}]\nonumber\\
    \label{1st_gen_X}
\end{align}
and obtain $Y_1$ and $Z_1$ by cyclic permutation of indices.
We can also express the $\{X_2,Y_2,Z_2\}$ variables as
\begin{align}
    X_2
    \approx (1 - \textbf{D}^4) X_{1}
    \label{2nd_gen_X}
\end{align}

The package \texttt{fastlisaresponse} performs TDI computations using the $\eta_{ij}$ variables, whereas \texttt{lisaglitch} injects glitches in terms of the $\text{tmi}_{ij}$ measurements. We therefore reformulate the standard TDI observables in terms of test-mass interferometer data.
By restricting glitch injections to the test-mass interferometers only, we set $\text{isi}_{ij} = 0$ and $\text{rfi}_{ij} = 0$, which simplifies the intermediary variables to
\begin{equation}
    \eta_{ij}
    =
    -\frac{1}{2}(\text{tmi}_{ij}+\textbf{D}\text{tmi}_{ji}),
    \end{equation}
which means that both generations of Michelson variables can be simplified by replacing all instances of $\eta_{ij}$ with the appropriate constant and delayed measurement $\text{tmi}_{ij}$ in~\cref{1st_gen_X,2nd_gen_X}.

For second generation Michelson variables, one obtains 

\begin{widetext}
\begin{align}
X_2 &= \frac{1}{2}(1 - \textbf{D}^4)^2(\text{tmi}_{12} - \text{tmi}_{13}) + \textbf{D}(1 - \textbf{D}^4)(1 - \textbf{D}^2)(\text{tmi}_{21} - \text{tmi}_{31})
\end{align}
\end{widetext}

We have algebraically checked the resulting expressions, including the final form for $X_2$, using the symbolic mathematics package \texttt{sympy}~\cite{10.7717/peerj-cs.103}.
Additionally, \href{https://lisa.pages.in2p3.fr/LDPG/wg6_inrep/pytdi/standard-combinations.html#x-1-y-1-z-1-as-functions-of-beatnotes}{existing documentation} from the \texttt{PyTDI} package details both the first- and second-generation variables in terms of base interferometer measurements, from which one can verify for themselves the derived expressions.

\section{Fisher matrix validation}\label{app_sec:FM_validation}

To assess the accuracy of our estimated FMs, we compare the FM-based posterior samples drawn from $\mathcal{N}(\boldsymbol{\theta_0},\boldsymbol{\Gamma}^{-1})$ with samples obtained from full MCMC analysis.

For each MCMC run, we adopted narrow uniform priors centered on the true parameters values to probe only the region around the global likelihood maximum.
This approach avoids the well-known multimodality that affects EMRI parameter estimation~\cite{chua_2022}. 
We employed the \texttt{eryn} sampler for 4,000 steps for the  prograde and strong-field EMRIs, and 1,700 steps for the retrograde EMRI, discarding the first half of each chain as burn-in to ensure convergence to the equilibrium distribution.

\begin{figure*}[p]
    \centering
    \includegraphics[width=\linewidth]{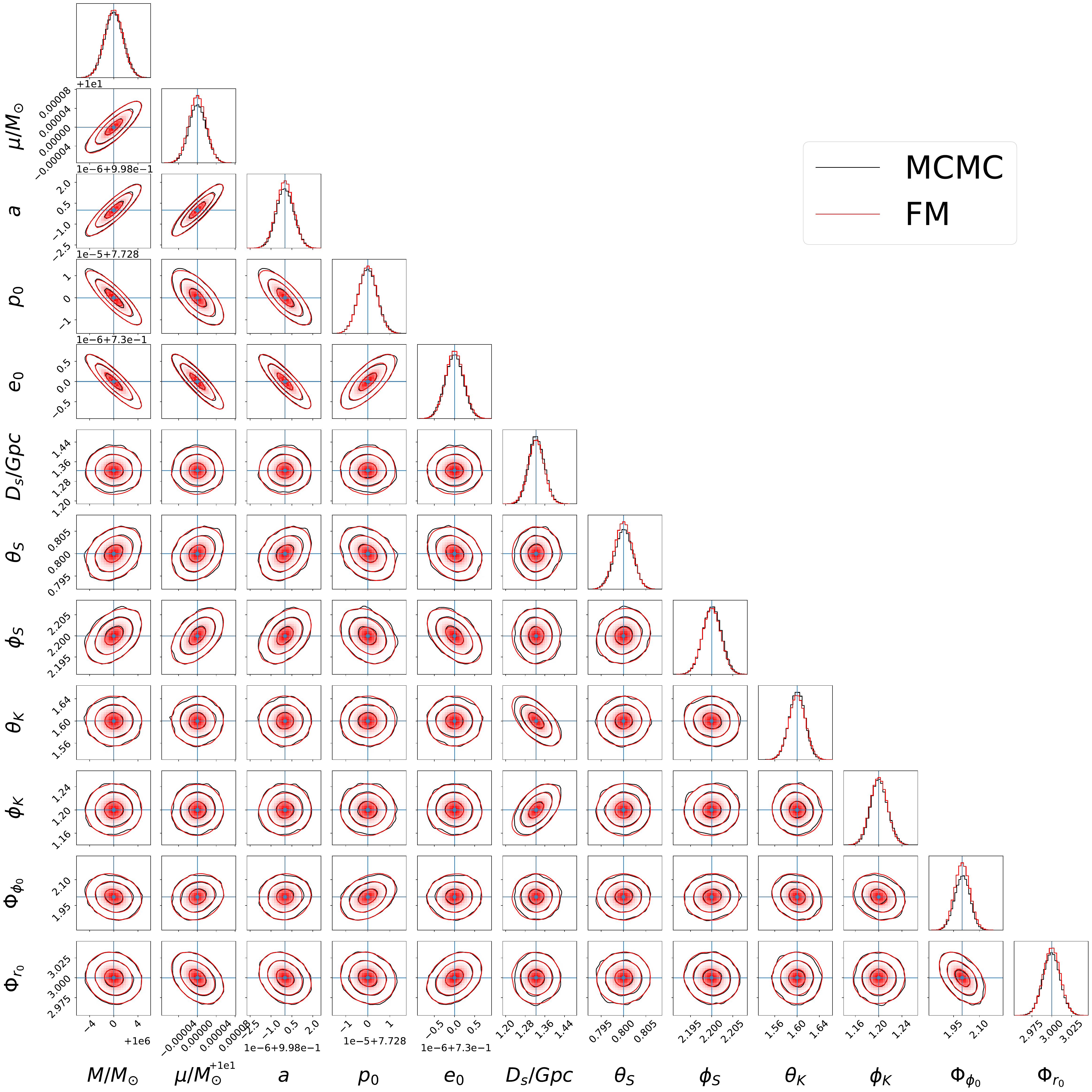}
    \caption{Marginal posteriors of the prograde EMRI as obtained by MCMC alongside the FM-derived posterior. The data contained only the EMRI signal in zero noise. The close agreement validates our Fisher matrix calculations.}
    \label{fig:Prograde_EMRI_FM_VS_MCMC_corner}
\end{figure*}

\begin{figure*}[p]
    \centering
    \includegraphics[width=\linewidth]{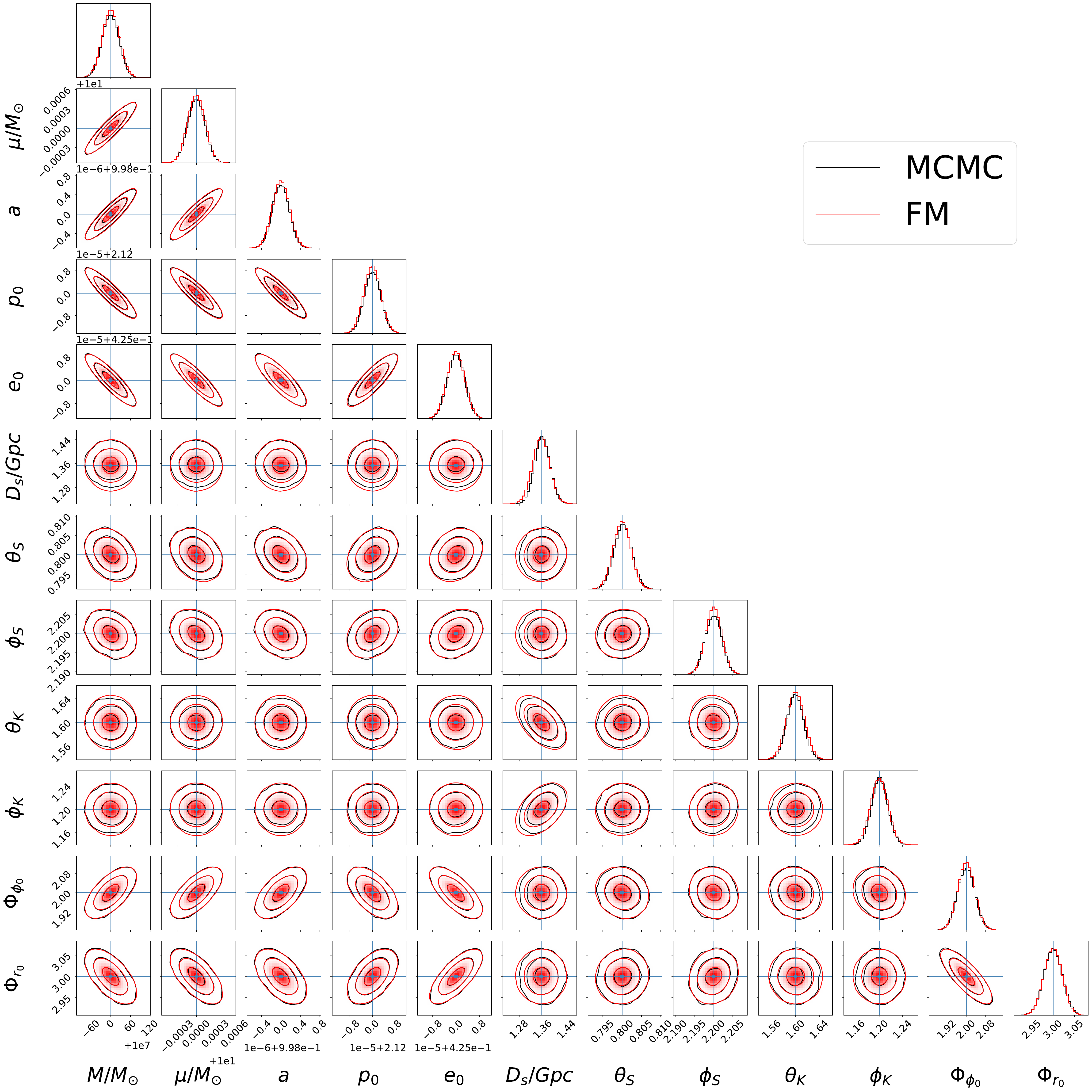}
    \caption{Marginal posteriors of the strong-field EMRI as obtained by MCMC alongside the FM-derived posterior. The data contained only the EMRI signal in zero noise. }
    \label{fig:Strongfield_EMRI_FM_VS_MCMC_corner}
\end{figure*}

\begin{figure*}[p]
    \centering
    \includegraphics[width=\linewidth]{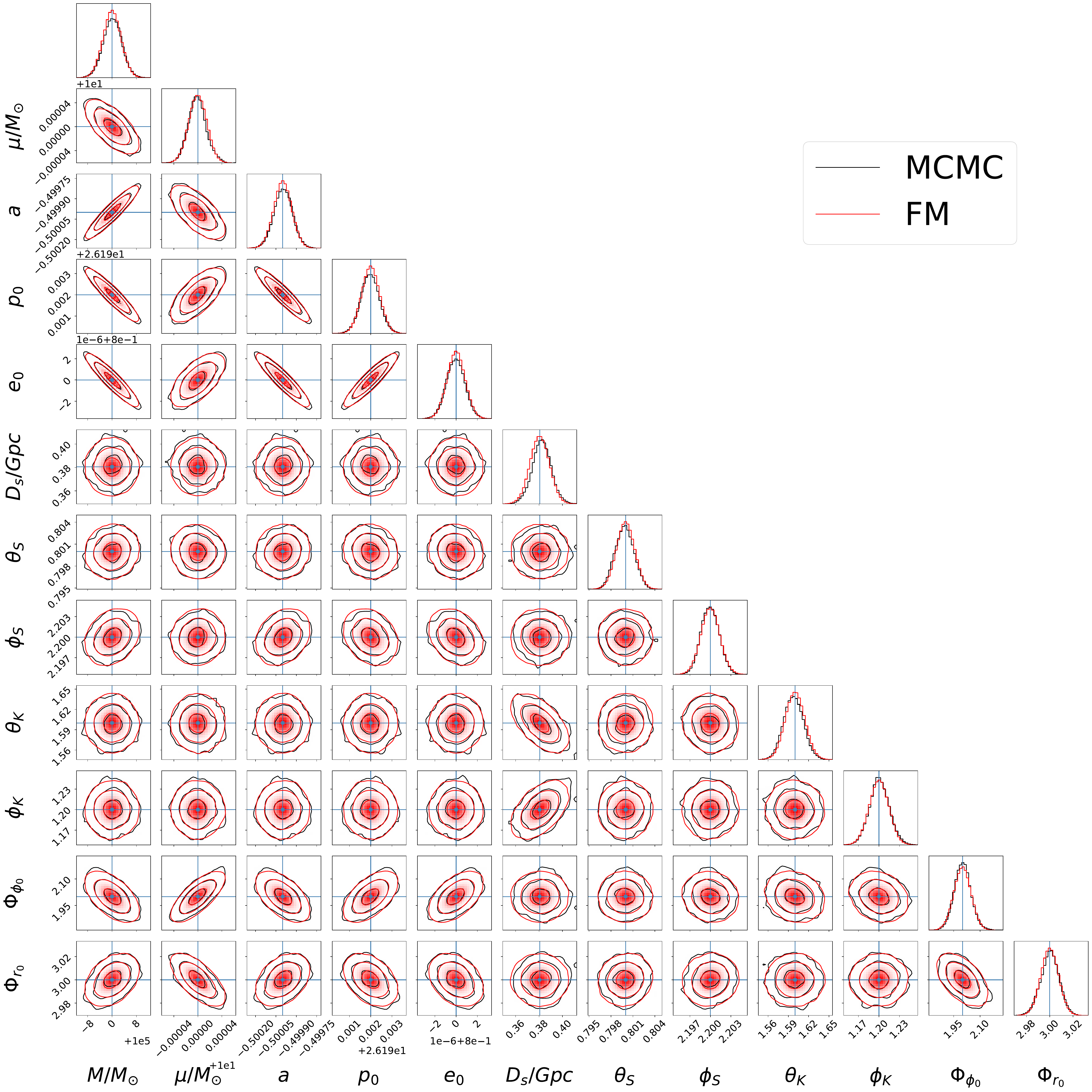}
    \caption{Marginal posteriors of the retrograde EMRI as obtained by MCMC alongside the FM-derived posterior. The data contained only the EMRI signal in zero noise. }
    \label{fig:Retrograde_EMRI_FM_VS_MCMC_corner}
\end{figure*}

Figures~\ref{fig:Prograde_EMRI_FM_VS_MCMC_corner}, \ref{fig:Retrograde_EMRI_FM_VS_MCMC_corner}, and \ref{fig:Strongfield_EMRI_FM_VS_MCMC_corner} visually demonstrate that despite the ill-conditioned nature of EMRI FMs~\cite{Vallisneri_2008}, the FM-derived posterior distributions closely match those obtained from MCMC sampling. 
This agreement indicates that our estimated FMs provide accurate local approximations to the likelihood, ensuring that the glitch-induced parameter fluctuations computed using Eq.~\ref{eq:glitch_error_eqn} are not compromised by FM inaccuracies.

To further validate Eq.~\ref{eq:glitch_error_eqn}, we compare a representative data realization's predicted glitch-induced parameter fluctuations with those estimated directly from MCMC analyses.
We find that the FM-derived parameter fluctuations agree closely with those from MCMC analyses.
For the prograde, retrograde, and strong-field EMRIs, the largest discrepancies are limited to $0.5\sigma$ for up to moderately mitigated cases.
For mitigation SNRs in excess of $400$, we find deviations of up to several sigma which implies that we are in a regime where large biases are unavoidable.
These comparisons demonstrate the accuracy of our estimated parameter fluctuations, with noticeable deviations only in the most nonlinear (high-SNR) regime.

\begin{figure*}[p]
    \centering
    \includegraphics[width=\linewidth]{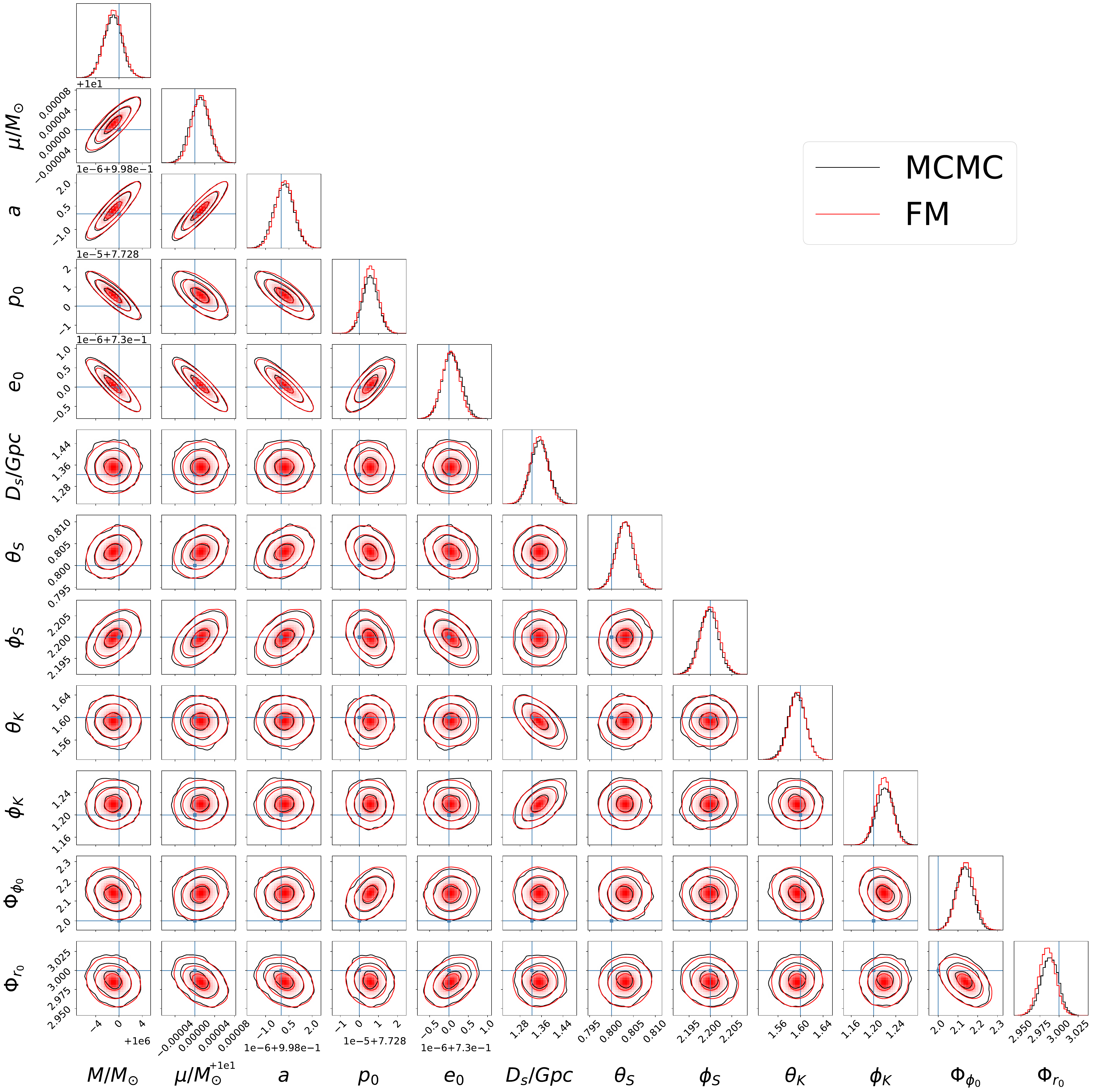}
    \caption{A comparison of the MCMC-derived and FM-derived posteriors for the prograde EMRI corrupted by a glitch background mitigated to SNR 400. We did not include stationary noise in the data.}
    \label{fig:Prograde_EMRI_bias_comparison}
\end{figure*}

% In addition to validating the accuracy of the FM itself, we also demonstrated the accuracy of FM-predicted posterior shifts by comparing with corresponding MCMC posteriors.
% These MCMC simulations were performed for each fiducial EMRI and a specific glitch background mitigated to the levels used in this study.
% We found that the shifts in our FM-predicted posteriors were in close agreement with MCMC, even for minimal glitch mitigation.
As a visual verification, Figure~\ref{fig:Prograde_EMRI_bias_comparison} shows matching biased posteriors for the prograde EMRI obscured by a particular glitch background mitigated to SNR 400.

\bibliographystyle{apsrev4-2}
\bibliography{refs}% Produces the bibliography via BibTeX.

\end{document}